\documentclass[aps,prl,twocolumn,showpacs,amsmath,amssymb]{revtex4-2}

\usepackage{graphicx}
\usepackage{amsmath}
\usepackage{setspace}
\usepackage{braket}
\usepackage{epstopdf}
\usepackage{comment}
\usepackage{hyperref}
\usepackage{bm}
\usepackage{xfrac}
\usepackage{xcolor}
\usepackage{subcaption}
\usepackage{floatrow}
\usepackage{mathrsfs}
\usepackage{chemformula}

\newcommand{\aap}{Astron. Astrophys.}

\newcommand{\jpb}{J. Phys. B. }
\newcommand{\quotes}[1]{``#1''}
\begin{document}

\title{Radiative electron attachment to rotating \ch{C3N} through dipole-bound states}

\author{Joshua Forer$^{1,2}$, Viatcheslav Kokoouline$^1$, Thierry Stoecklin$^2$}
\affiliation{$^1$Department of Physics, University of Central Florida, Orlando, Florida 32816, USA\\
$^2$Institut des Sciences Mol\'eculaires, Universit\'e de Bordeaux, CNRS UMR 5255, 33405 Talence Cedex, France}

\date{\today}

\begin{abstract}
  The role of a large dipole moment in rotating neutral molecules interacting with low-energy electrons is studied using an accurate {\it ab initio} approach accounting for electronic and rotational degrees of freedom.
  It is found that theory can reproduce weakly-bound (dipole-bound) states observed in a recent photodetachment experiment with \ch{C3N-} [Phys. Rev. Lett. {\bf 127}, 043001 (2021)].
  Using a similar level of theory, the cross section for radiative electron attachment to the \ch{C3N} molecule, forming the dipole-bound states, was determined.
  The obtained cross section is too small to explain the formation of \ch{C3N-} in the interstellar medium, suggesting that it is likely formed by a different process.
\end{abstract}

\maketitle


{\it Introduction}
Despite the large amount of molecules detected in interstellar and circumstellar environments, only six molecular anions have been detected~\cite{C3Nm-Thaddeus-2008,C4Hm-Cernicharo2007,C5Nm-Cernicharo2008,C6Hm-McCarthy2006,C8Hm-Bruenken2007}: \ch{CN-}, \ch{C3N-}, \ch{C5N-}, \ch{C4H-}, \ch{C6H-}, and \ch{C8H-}. Their formation is not yet well understood. Initially, it was proposed that such anions are formed by the process of radiative electron attachment (REA), in which  a free electron can attach to the neutral molecule $M$ and  release excess energy via photon emission
\begin{equation}
M + e^- \to M^- + \hbar\omega\,.
\end{equation}
However, previous calculations \cite{khamesian16,Tennyson-C3N-JPB-2011,Miguel-REA-PRA-2019.99.033412,Miguel-Carbon-Chain-ESC-2019} have shown that REA rates are too low to explain the observed abundance of \ch{CN-}, \ch{C3N-}, and \ch{C5N-}.
It was then suggested that the REA mechanism could take place via a \quotes{doorway} weakly bound dipole state (DBS) \cite{DBS-energy-C3Nm-PRL-2021}, first described by Fermi and Teller in 1947 \cite{critical-dipole-Fermi-Teller-1947-PhysRev.72.399}.
For a point dipole interacting with a charge, the existence of an infinite number of  DBSs was predicted  for dipole values larger than a critical value of approximately 0.65 a.u. \cite{critical-dipole-Crawford_1967} (a.u. here and below are atomic units).
Including additional molecular effects, however, modifies this value while reducing the number of DBSs to just a few.
Other contributions to the long-range potential being associated with the electrostatic and polarization potentials need  to be considered when modeling a neutral polar molecule interacting with a point charge and, for these reasons, some authors have also investigated the possible existence of quadrupole and polarization bound states \cite{Desfrancois-PRL-1994-DBS-dipole,Desfrancois-DBS-1998}.
Experimentally, the minimum dipole moment required for the formation of stable dipole-bound anions of common closed-shell molecules has been measured to range between 2-2.5 D.  Over the years,  the effect of vibration \cite{Couplage-vib-DBS-PRL-2020-124.203401} and rotation of the dipolar molecule  on the DBS was also investigated \cite{Brinkman-DBS-rotation-JCP-1993,Brauman-rotation-dbs-lifetime-JPCA-2005}, showing that the critical moment for a rotating dipole is larger than that of a stationary dipole.
However, a quantum study of REA through a dipole-bound state including both the rotation of the molecule and the short range part of the electron-molecule potential  is missing.
In this study we consider the REA process via the DBSs of one of the anions  detected in the ISM:  \ch{C3N-}.
This molecule is a good candidate because the \ch{C3N} dipole is  supercritical and the existence of a \ch{C3N-} DBS  was  demonstrated in a recent ion trap experiment.  \cite{DBS-energy-C3Nm-PRL-2021}

{\it Theoretical model}
The theoretical model of the present study is based on wave functions (bound and continuum states) of the \ch{e-}-\,\ch{C3N} system obtained by solving the Schr\"odinger equation. The potential energy operator of the \ch{e-}-\,\ch{C3N} system is calculated in the molecular center-of-mass frame with the $\hat{z}$-axis directed along the molecular axis, pointing towards the N atom, while the incident electron's polar coordinates in this frame are denoted by $(r,\theta) $.
To highlight the effect of the short range contribution, two different potentials will be used.
The first potential, $V_\text{dip}$, is restricted to only the long-range interaction between a point charge and a dipole
\begin{equation}
	V_\text{dip}=-\dfrac{\mu e}{r^{2}}P_{1}(\cos \theta)\,,
	\end{equation}
where $ \mu $ is the dipole moment of \ch{C3N} ($\mu=$1.3 a.u.) and $ P_{1}(\cos \theta) $ is a Legendre polynomial.

The second interaction potential  $ V_\text{ai} $ between neutral C$ _{3} $N and the impinging electron is  an \textit{ab initio} potential made of four contributions and obtained as a function of the electronic wave function $ \psi^{e} $ of  C$_{3} $N
\begin{equation}\label{eq:Vai}
	V_\text{ai}(\psi^{e})=V^{e}_\text{Stat}(\psi^{e})+V_\text{nuc}+V_\text{Ex}^\text{HFEGE}(\psi^{e})+V_\text{copol}(\psi^{e})\,,
\end{equation}
where $ V^{e}_{Stat}(\psi^{e})$ is the electronic part of  the static potential,  $ V_{nuc } $ is the contribution of the nuclear potential, and the two remaining terms are exchange and correlation-polarization contributions, respectively.
The latter two contributions are obtained in the usual density-functional forms  of Morrison and Collins \cite{HFEGE-Morrison-1978-PRA} and Padial and Norcross \cite{Padial-Norcross-1984}.
The perpendicular $\alpha_\perp $ and parallel $\alpha_\| $ polarizabilities, and  the ionization potential of \ch{C3N}, needed for the potential of Eq.(\ref{eq:Vai}), are 27.11, 60.35, and 0.53 a.u., respectively.
The ground electronic state ($\ch{X} {}^2\Sigma^+$) wave function of \ch{C3N}, $ \psi^{e}$, is calculated for its equilibrium geometry.
At equilibrium, the molecule is linear with the following three internuclear distances in the C-C-C-N arrangement: 2.30, 2.62, and 2.13 a.u., respectively (see the supplementary material of Ref.~\cite{Miguel-Carbon-Chain-ESC-2019}).
The same geometry is used within the fixed nuclei approximation for performing the \textit{ab initio} and REA calculations  as well as those of the \ch{C3N-}$ (\ch{X} {}^1\Sigma^+) $ bound states within the rigid rotor approximation. The \textit{ab initio} calculations are performed with the {\tt MOLPRO} \cite{Molpro2012} software suite at the MCSCF level and using the {\tt aug-ccpV6Z(d)} atomic orbital basis~\cite{Dunning-aug-ccpv6Z-1996}.
The obtained dipole moment of C$ _{3} $N of 1.3 a.u. is in very good agreement with the CAS-CI value 1.4 a.u. by \citet{C3N-dipole-CAS-CI-Harrison_2011}, suggesting that the main contribution to the long-range \ch{e-}-\ch{C3N} potential is well described at this level of theory.

In order to obtain the potential as a series of Legendre polynomials
\begin{equation}
	V_\text{ai}=\sum\limits_{\lambda=0,80}  C_{\lambda}(r)P_{\lambda}(\cos \theta)\,
\end{equation}
the \textit{ab initio} multi-centered gaussian basis set is first analytically shifted and expanded around the centre of mass in a  symmetrized, real spherical harmonics basis set  \cite{spherical-harmonics-expansion-gaussian-1994} for values of $l$ up to 380. From this development, analytical Legendre polynomial expansion-coefficients of the electronic potential {\color{black}$ V^{e}_\text{Stat}(\psi^{e})$ --- by far the strongest contribution ---} are obtained using the standard procedure  \cite{Stoecklin-C2H2-1994}.

\begin{figure}
	\centering
  \includegraphics[page=1,width=0.9\textwidth]{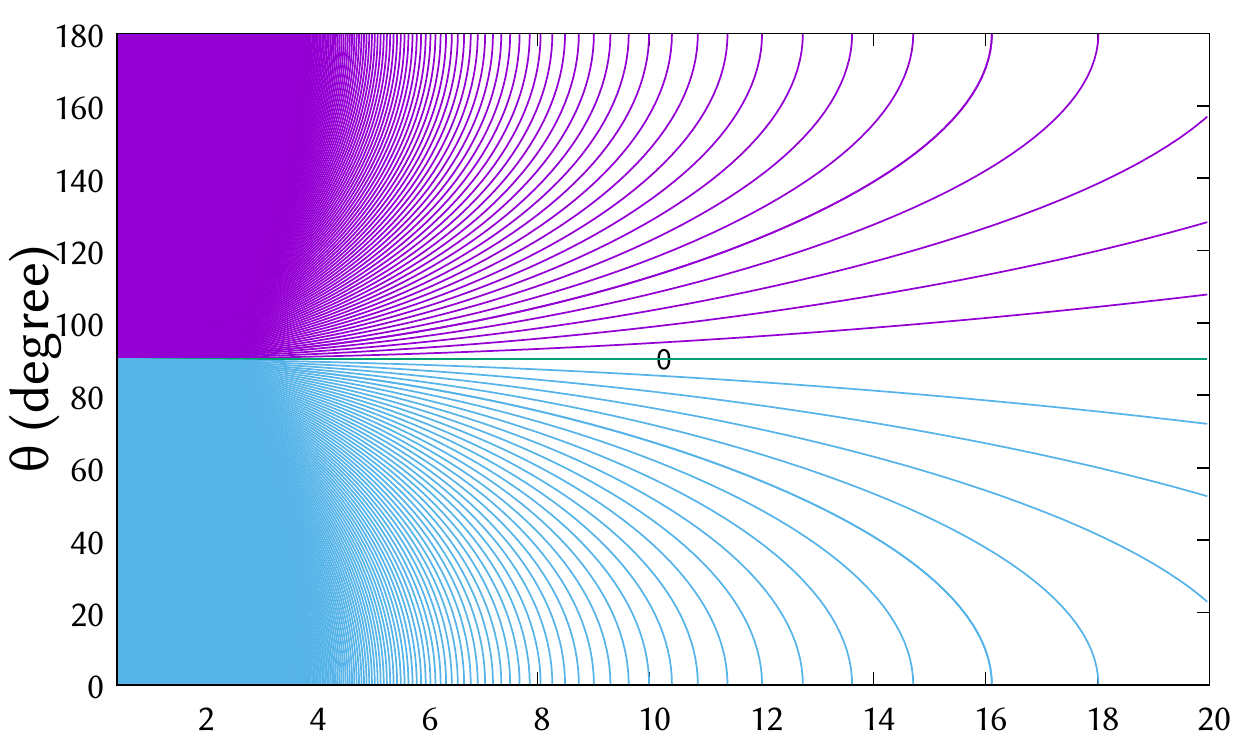}
  \includegraphics[page=2,width=0.9\textwidth]{fig1.pdf}
  \caption{The two potentials used in the study,  $ V_\text{dip}$ (upper panel) and $V_\text{ai}$ (lower panel),  shown as a function of the Jacobi coordinates $r$ and $\theta$. 	The energy-dependent potential $V_\text{ai}$ is represented at an electron scattering energy -0.1 eV ($\sim$\color{black} -806.5~cm$^{-1}$).
  {\color{black}The origin of the figure is the C$ _{3} $N center of mass. The positions of the four atoms  are also reported.}}
	\label{fig:pot}
\end{figure}

Figure \ref{fig:pot}  compares the two potentials. The potentials are significantly different at small distances, demonstrating that the point dipole potential may not accurately describe the \ch{C3N} DBS.
The $ V_\text{ai} $ potential accounts for the short- and long-range contributions.
The nuclear potential $ V_{nuc} $ is strongly attractive near the nuclei where it dominates all the other contributions.
For larger, but not very large, values of $r$, the sum of the repulsive electronic potential and the attractive exchange and correlation-polarization contributions makes the overall potential less attractive than the pure charge-dipole potential $V_\text{dip}$.
We do not expect high-energy continuum states and resonances to be accurately represented by the  {\color{black}$V_\text{ai}$ } potential, but it should describe the process well, qualitatively, at low collision energies where {\color{black}the balance between } the  long- and mid-range potential controls the dynamics.

{\it Dipole-bound states}
The close-coupling method is used to solve the Schr\"odinger equation for bound and continuum states using the Magnus propagator within the Newmat code as detailed in a previous study \cite{He-MnH-Bound-states}.
The calculations included 40 rotational states of \ch{C3N} and were propagated from $r=0.1$ to 300 a.u. while the  boundary between propagation and counter-propagation was fixed at $3.2$ a.u. DBS energies obtained for the two potentials and different values of total angular momentum $J$ are given in Table	\ref{table_bound_E}.

\begin{table}[h]
	\caption{DBS energies (cm$^{-1}$) obtained for the two potentials for different values of $J$.}
	\begin{tabular}{c c c}
		\hline
		$J$ \quad &  $E(V_\text{ai})$  & $E(V_\text{dip})$ \\ \hline
		0   \quad & -6.87              & -22.13            \\
		1   \quad & -6.57              & -21.80            \\
		2   \quad & -5.87              & -21.14            \\
		3   \quad & -4.90              & -20.15            \\
		4   \quad & -3.51              & -18.84            \\
		5   \quad & -1.87              & -17.19            \\
		6   \quad &                    & -15.21            \\
		7   \quad &                    & -12.91            \\
		8   \quad &                    & -10.27            \\
		9   \quad &                    & -7.30             \\
		10  \quad &                    & -4.01             \\
		11  \quad &                    & -0.38             \\
		\hline
	\end{tabular}
	\label{table_bound_E}
\end{table}

The potential $V_\text{dip}$ has bound states up to $J=11$, while $V_\text{ai}$ only supports bound states up to $J=5$.
This difference results from the less attractive mid-range part of $V_\text{ai} $ compared to $V_\text{dip}$.
In both cases, the only possible states with $J>0$ have odd parity if $J$ is even, and even parity if $J$ is odd.
The spacing between successive levels in both cases follows the usual law for a linear molecule with a rotational constant of \ch{C3N-}, which is very close to the one of \ch{C3N} {\color{black}(0.16504 cm$ ^{-1} $ \cite{Gottlieb1983}).}
{\color{black} The $J=0$ bound-state energy obtained with the pure charge-dipole potential $V_\text{dip}$ is quite large when compared to the experimental value $ (-2\pm 1$ cm$^{-1})$ of Wester \cite{DBS-energy-C3Nm-PRL-2021}. The corresponding value for $V_\text{ai}$  is closer to the experiment but still three times larger.  If we take into account (1) the experimental temperature of the ion trap  (16 K), which allows populating several rotationally-excited dipole bound states, and (2) the energies computed with $V_\text{ai}$, we obtain a Boltzmann-average peak position in the photodetachment spectrum of $-3.8$~cm$^{-1}$, which is in good agreement with its experimental estimate. The agreement within a few cm$^{-1}$ with the theoretical value,  obtained neglecting rotation  by \citet{Gianturco-DB-C3Nm}  (-2~cm$^{-1}$)  from an MRCI calculation of \ch{C3N-},  is also satisfactory and demonstrates the validity of the $V_\text{ai}$ potential.  }

\begin{figure}
	\centering
	\includegraphics[width=0.9\textwidth]{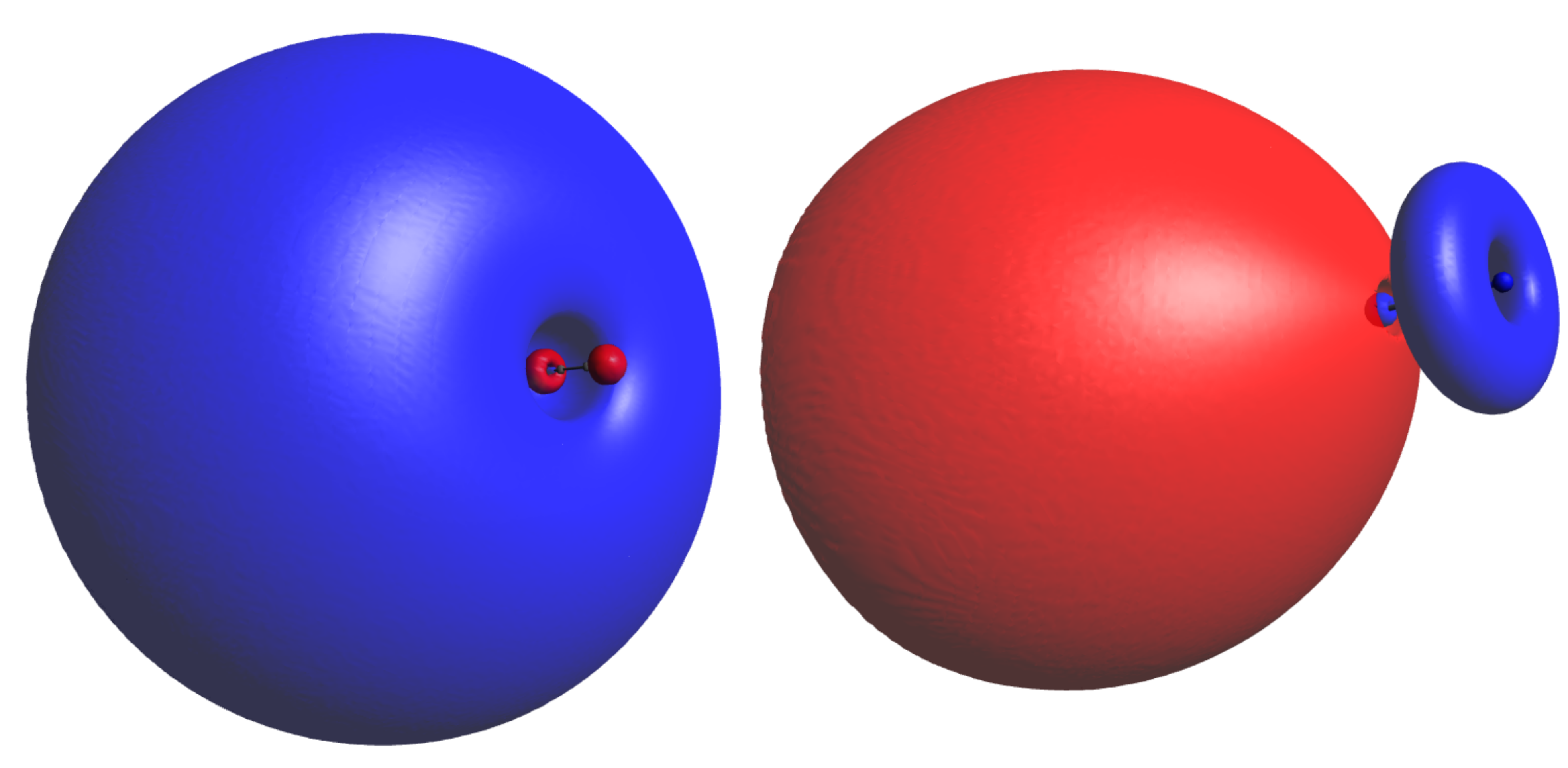}
	\caption{Dipole-bound states of \ch{C3N-} for $J=0$ with energy -6.87~cm$^{-1}$(left) and  $J=1$ with energy -6.57~cm$^{-1}$(right)}
	\label{fig:wf}
\end{figure}

Figure \ref{fig:wf} shows two examples of DBS wave functions obtained for the $V_\text{ai}$ potential for $J=0$ with energy {\color{black}$-6.87$ }cm$^{-1}$ and $J=1$ with energy {\color{black} $-6.57$ cm$^{-1}$}. The DBS wave functions with higher $J$ qualitatively look the same but have additional nodes along the $\hat{z}$-axis.
Additionally, wave functions obtained with the $V_\text{dip}$ potential have the same behavior as the ones obtained with $V_\text{ai}$ at large values of $r$, but have fewer features with small amplitudes at short distances due to the simpler behavior of $V_\text{dip}$ in this region (see Fig.~\ref{fig:pot}).

{\it Electron-molecule scattering and radiative attachment}
For REA calculations, \ch{e-}-\,\ch{C3N} continuum-state wave functions are also needed.  A key difference between  our approach and that of Lara-Moreno \textit{et al.} \cite{Miguel-REA-PRA-2019.99.033412} is that the effect of rotation on the REA process is accounted for. The space-fixed frame is then more appropriate than the body-fixed frame used in that previous study. We furthermore take advantage of using a local model of the \ch{e-}-\,\ch{C3N} potential and  use a method initially developed for radiative association \cite{ayouz11,Stoecklin-RA-2013} and later adapted to REA  \cite{douguet13,douguet15}.

We do not resolve molecular vibration and write the rigid rotor scattering wave function of the system  in Jacobi coordinates  in the following  form
\begin{equation}
	\psi^i(\vec{r},\hat{Z}) = \frac{1}{r} \sum\limits_{jl} \chi^{JM}_{jl} (r) \; Y^{JM}_{jl}(\hat{r},\hat{Z})
	\label{psi_initial}\,,
\end{equation}
where $ \vec{r} $ and $ \hat{Z} $ are respectively  the electron coordinate vector  and the orientation angles of the \ch{C3N} axis in the space-fixed frame. The indices $j$ and $l$ are, respectively, the \ch{C3N} rotational  and electron  orbital  quantum numbers, while the total angular momentum of the system is $\vec{J} = \vec{j} + \vec{l}$.
The $ Y^{JM}_{jl}(\hat{R},\hat{r}) $ angular  functions take the usual form
\begin{equation}
 Y^{JM}_{jl}(\hat{r},\hat{Z}) = \sum\limits_{m_jm_l} \braket{jm_jlm_l | JM} Y^{m_l}_{l}(\hat{r}) Y^{m_j}_j(\hat{Z})\,,
 \label{angular_basis}
\end{equation}
where $M$, $m_j$, and $m_l$ are, respectively, the projections of $\vec{J}$, $\vec{j}$, and $\vec{l}$ on the space-fixed $\hat{z}$-axis.

For a given value of $J$, the radial part of the scattering  wave function $ \chi^{J}_{jl} (r)$  is the solution of the driven differential equation \cite{Stoecklin-RA-2013}
\begin{eqnarray}
	\left(
		\frac{d^2}{dr^2} - \frac{l(l+1)}{r^2} + k_j^2(E) - 2 m_r U_{j'l',jl}^{J}(r)
	\right)\nonumber
	\chi_{j'l',jl}^{J} (r) \\
	=\lambda_{jl}^{\alpha J}(r)\,,
	\label{cc_lambda}
\end{eqnarray}
where $U_{j'l',jl}^{J}(r)$ is the potential matrix in  basis set (\ref{angular_basis}). The right-hand-side term  is called
the driving term for a given initial state of the \ch{e-}-\,\ch{C3N} system and final state of $\ch{C3N-}$, each respectively characterized by the sets of quantum numbers $ (j,l) $ and $ (\alpha, J')$.
It is real-valued and results from the dipolar coupling of the initial (scattering) and final (bound) states within the dipolar approximation.
It is given by
\begin{equation}
	\lambda_{jl}^{\alpha J}(r) = -2m_r \int d\hat{r} \; d\hat{Z} \; Y^{JM}_{jl}(\hat{r},\hat{Z}) \mu(\vec{r},\hat{Z}) \psi^f_{\alpha J'}(\vec{r},\hat{Z})\,,
\end{equation}
 where $m_r$ is the reduced mass of the system. {\color{black} The final DBS functions $ \psi^f_{\alpha J'}(\vec{r},\hat{Z})$, presented in the previous section, were also calculated  in the space fixed frame and expanded in the basis given by equation (\ref{angular_basis}), i.e.,}
\begin{equation}
	\psi^f_{\alpha J'}(\vec{r},\hat{Z}) = \frac{1}{r} \sum\limits_{jl} \omega^{J'M'}_{jl}(r) \; Y^{J'M'}_{jl}(\hat{r},\hat{Z})\,.
	\label{psi_final}
\end{equation}

The two sets of DBSs  obtained using the potentials $V_\text{ai}$ and $V_\text{dip}$ are then used to calculate the REA cross section with a rotational basis of 40 states.
The REA cross section is given by the following equation
\begin{equation}
	\sigma^\text{REA}_f =
	\frac{g_a}{g_n} \frac{8\pi^2}{3k_e^2c^3}
	\sum\limits_{J,J',\alpha} \omega_{\alpha}^3 |M_{j,J}^{\alpha J'} |^{2}\,,
	\label{eq:sigma}
\end{equation}
where the transition dipole moment $ M_{j,J}^{\alpha J'}  $ is obtained from the driving term \cite{Stoecklin-RA-2013}, $g_a$ and $g_n$ are respectively the  degeneracy factors of the anion and the neutral, and $\omega_{\alpha}$  is the frequency of the emitted photon.

\begin{figure}
	\centering
	\includegraphics[width=0.9\textwidth]{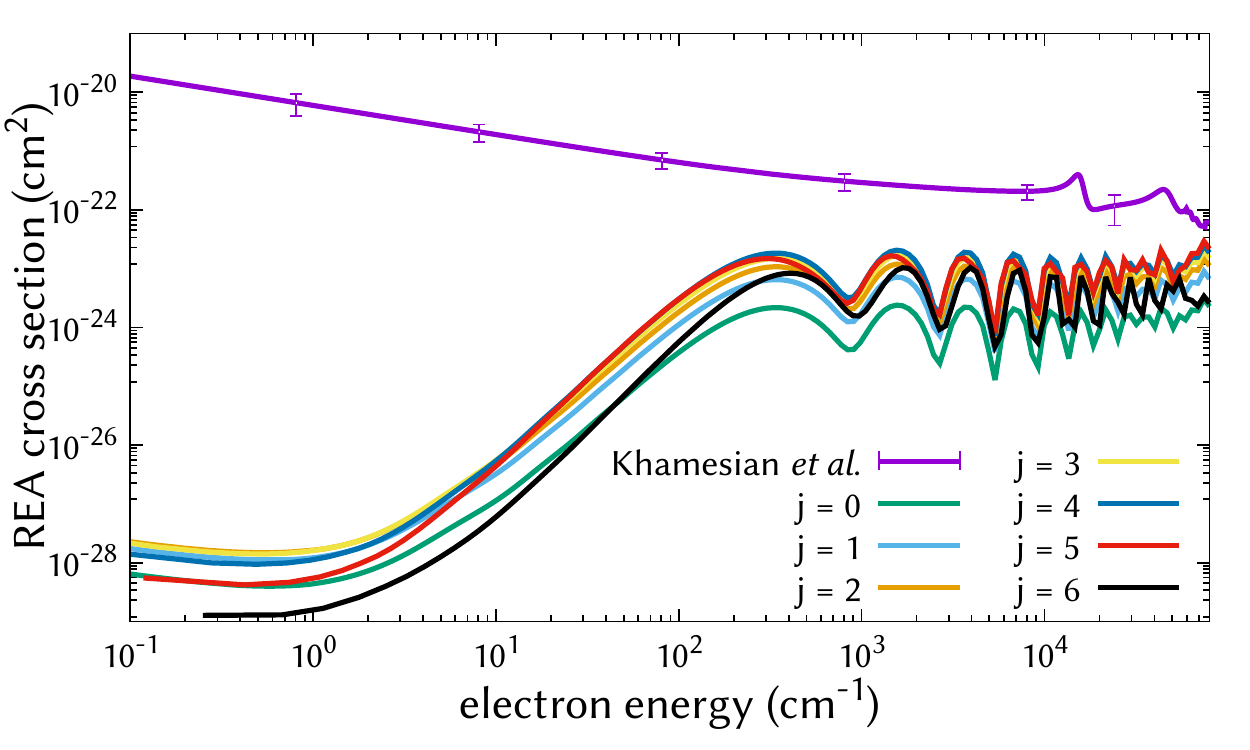}
	\caption{
    The REA cross sections for different initial rotational states $j$ of \ch{C3N-} using the $V_\text{ai}$ potential.
    The figure shows the result (purple curve with error bars) of the previous REA study \cite{khamesian16}.
    The error bars on the curve represent the uncertainty of the model used in that study.
  }
	\label{fig:REA_XS}
\end{figure}

{\it Results}
{\color{black} Figure \ref{fig:REA_XS} shows the REA cross section  for different initial rotational states $j$ of \ch{C3N} using the $V_\text{ai} $ potential. The cross section is summed over all possible final DBS states, i.e. over different $J$. The cross section for $j=0$ is 2-4 times smaller than for larger values of $j$. At very small energies, below 1 cm$ ^{-1} $, the cross section is inversely proportional to the collision energy, as expected from Eq.~(\ref{eq:sigma}). Between 1 and 100 cm$ ^{-1}$, it increases rapidly as, approximately, the square of the collision energy due to the $\omega_\alpha^3/k_e^2$ factor in Eq.~(\ref{eq:sigma}). Above  100~cm$ ^{-1} $, the cross section oscillates due to the same behaviour of the transition dipole moment in Eq.~(\ref{eq:sigma}), which varies due to the change in position of nodes in the initial continuum wave functions with the energy.}

The figure also shows the result of the previous REA study  \cite{khamesian16}, in which the REA cross section for transitions to the ground electronic state of \ch{C3N-} were considered. At energies below 1~eV, the present cross sections are much smaller than the one in Ref.~\cite{khamesian16}.
This is expected for low collision energies because the overall magnitude of the REA cross sections is governed by the $\omega_\alpha^3$ factor in Eq.~(\ref{eq:sigma}), which is much smaller for transitions to DBSs than to ground electronic state of \ch{C3N-}.
The \ch{C3N} affinity is $\sim$34727~cm$^{-1}$ \cite{DBS-energy-C3Nm-PRL-2021}.
The cross sections approach the one obtained by \citet{khamesian16}, approximately, only at scattering energies that are larger than the \ch{C3N-} affinity.

In their study, \citet{khamesian16} concluded that the obtained REA cross section is too small to explain the observed abundance of \ch{C3N-} ions in the ISM by formation via REA. In the present study, REA cross sections are even (significantly!) smaller at energies relevant for the interstellar medium.
This means that the idea \cite{carelli14} that DBSs can increase the REA cross section and possibly explain the observed abundance of \ch{C3N-} by REA to \ch{C3N} should be abandoned.

{\it Concluding}, we would like to stress the following findings of the present study.
Energies of dipole-bound states of rotating \ch{C3N-} molecular ion were obtained using an accurate computational approach. In total, six weakly-bound states were found for $J=0-5$. Cross sections for radiative attachment to the \ch{C3N} molecule forming the weakly-bound states of \ch{C3N-} were obtained for several initial rotational states $j=0-6$ of the neutral molecule and for a large interval of collision energies. The obtained REA cross sections are significantly smaller than those previously obtained for the process towards the ground electronic state of \ch{C3N-}, which was, in principle, expected.
\begin{itemize}
  \item It was concluded that the  \ch{C3N-} abundance observed in the ISM cannot be explained by formation of these ions by REA towards weakly or deeply bound electronic states of the ion. {\color{black}The present results obtained at the rigid rotor level suggest that the  effect of the formation of long-lived resonances via vibrationally excited bending vibrations should be considered in future studies. The alternative formation mechanism of large molecular anions via reactions of  carbon anions with N atoms should also be considered for  C$_3$N$^-$  as suggested by  \citet{bierbaum_2011} and  \citet{CNm-Agundez2010}. Five other negative ions observed in the ISM: \ch{CN-}, \ch{C5N-}, \ch{C4H-}, \ch{C6H-}, and \ch{C8H-}, deserve separate studies of this kind, as their formation mechanisms are probably not the same.   }
  \item A technical result, imporant for theorists: we have found that the short-range part of the electron-molecule interaction significantly influences the REA cross sections for energies below a few eV.
  \item The present approach can be applied to study weakly-bound states and electron scattering for systems where the strong quadrupole interaction is dominant at large distances, such as \ch{TCNB-} \cite{liu2020observation}.
\end{itemize}

\section{Acknowledgements}
This work acknowledges the support from the National Science Foundation, Grant No.2110279, and the Fulbright-University of Bordeaux Doctoral Research Award.


%

\end{document}